\documentclass[12pt]{iopart}

\usepackage{iopams}  
\usepackage{graphicx}
\usepackage{stix}
\usepackage{xcolor}
\usepackage{upgreek}
\begin{document}

\title[Mean transition path velocities of colloidal particles surmounting energy barriers]{Experimental measurement of mean transition path velocities of colloidal particles surmounting energy barriers}

\author{Brandon R. Ferrer$^{1}$, Juan Ruben Gomez-Solano$^{1,*}$}

\address{$^{1}$Instituto de F\'isica, Universidad Nacional Aut\'onoma de M\'exico, Ciudad de M\'exico, C\'odigo Postal 04510, Mexico,}

\ead{$^*$ r\_gomez@fisica.unam.mx}
\vspace{10pt}
\begin{indented}
\item[]February 2024
\end{indented}

\begin{abstract}
Transition paths are rare events occurring when a system, thanks to the effect of fluctuations, crosses successfully from one stable state to another by surmounting an energy barrier. Even though they are of great significance in many mesoscale processes, their direct determination is often challenging due to their short duration as compared to other relevant time-scales. Here, we measure the local average velocity along transition paths of a colloidal bead embedded in a glycerol/water mixture that hops over a barrier separating two optical potential wells. Owing to the slow dynamics of the bead in this viscous medium, we can spatially resolve the mean velocity profiles of the transition paths for distinct potentials, which agree with theoretical predictions of a model for the motion of a Brownian particle traversing a parabolic barrier. This allows us to experimentally verify various expressions linking the behavior of such mean velocities with equilibrium and transition path position distributions, mean transition-path times and mean escape times from the wells. We also show that artifacts in the mean velocity profiles arise when reducing the experimental time resolution, thus highlighting the importance of the sampling rate in the characterization of the transition path dynamics. Our results confirm that mean transition path velocity establishes a fundamental relationship between mean transition path times and equilibrium rates in thermally activated processes of small-scaled systems.
\end{abstract}

%
\vspace{2pc}

%
%
%

\section{Introduction}\label{sect:intro}

Thermally activated transitions take place in diverse mesoscopic systems that exhibit multistability, such as in biomolecular folding reactions~\cite{woodside2006,chung2009}, protein association~\cite{sturzenegger2018,kim2018}, adatom diffusion~\cite{Jamnig2019,alghannam2017}, nanomagnetic information storage~\cite{coffey2012,Hong2016}, particle hopping in dense colloidal suspensions \cite{schweizer2003,ma2019}, self-feedback in cavity optomechanical systems~\cite{liu2021}, etc. In such processes, thermal fluctuations boost the escape of the system from a particular stable state by providing it with the necessary energy to surmount the corresponding barrier, thus allowing it to explore other local minima in its energy landscape. In general, they involve two characteristic time-scales that depend on the coupling of the system with its environment, thus encoding different pieces of information on the microscopic mechanisms underlying  the transition. The first one is the mean escape time, \textit{i.e.} the time that the system dwells on average in a state corresponding to a certain potential minimum until it fully escapes, whose inverse represents the transition rate of the activated process. The second one is the mean duration of transition paths, which are transient parts of the trajectory of a coordinate characterizing the dynamics of the system as it spontaneously traverses from a specific location on the barrier to another on the other side without recrossing the former. Since they actually correspond to successful crossing events over the barrier, their average duration is significantly shorter than the mean escape time, the latter being a mean first passage time that encompasses all failed attempts before a transition path occurs.

Much of our current understanding on thermally activated processes has been gained thanks to the formulation of Kramers' escape problem, which models them as the diffusive dynamics of a particle in contact with a heat bath that starts in a potential well and eventually hops over a barrier~\cite{kramers1940}. Since then, numerous theoretical efforts have been devoted to the determination of mean escape times and transition rates for diverse systems under various physical conditions~\cite{grote1981,hanggi1982,carmeli_1983,carmeli1983,munakata1985,melnikov1986,talkner1988,pollak1989,haenggi1990,melnikov1991,berezhkovskii1997,metzler2000,filliger2007,zhou2014,kappler2018,schuller2020,cherayil2022,wang2023,kumar2024}. Many of those predictions have been experimentally validated by direct visualization of colloidal beads suspended in fluids such as water~\cite{Simon1992,Lowell1999,Cohen2005,Ma2013,Su2017,yoon2020,Chupeau2020,Zijlstra2020,lyons2024}, gases~\cite{Rondin2017} and viscoelastic liquids~\cite{Ferrer2021}, hopping across potential wells sculpted by, \textit{e.g.} optical, electrophoretic and gravitational methods, which mimic the energy landscape considered in Kramers' problem. Along the same lines, a number of theoretical results have been derived over the past two decades on the statistical properties of transition paths of Brownian systems surmounting energetic barriers~\cite{hummer2004,zhang2007,chaudhury2010,malinin2010,kim2015,makarov2015,pollak2016,laleman2017,janakiraman2018,berezhkovskii2018,medina2018,berezhkovskii2019,caraglio2020,li2020,satija2020,singh2021,dutta2021,goswami2023}, which can shed light on the intricate molecular mechanism responsible for equilibrium reactions in condensed matter phases. As a matter of fact, in the last few years it has been possible to experimentally detect transition paths in protein and nucleic acids folding reactions thanks to recent improvements in single-molecule techniques~\cite{neupane2012,chung2012,yu2012,
chung2013,lannon2013,chung2014,chung2015,truex2015,neupane2016,neupane2018,neupane_2018,hoffer2019,kim2020,tripathi2022}. The results of such experiments provide clear evidence that the probability distributions of transition-path times exhibit the typical non-exponential behavior with an asymmetric peak derived from diffusive models. Nevertheless, to the best of our knowledge there is only one experimental work that has quantitatively verified the predicted dependence of the transition-path time distribution for diffusive dynamics on the parameters of a parabolic barrier by monitoring the motion of a colloidal bead in water through a bistable optical trap \cite{Zijlstra2020}. Measuring other features of transition paths analyzed in theoretical models, \textit{e.g.} their average shape and probability distributions of reaction coordinates, is more challenging due to experimental limitations of the spatio-temporal resolution within the transition region over the barrier, where multiple reversals of the dynamics occur before reaching the target. Indeed, as shown by recent numerical studies, artifacts due to an insufficient time resolution can emerge in the determination of statistical properties of transition paths over a barrier \cite{makarov2022}, even in simple instances of free diffusion \cite{song2023}.

It is worth pointing out that an important quantity that statistically describes the barrier-crossing process of a system transiting between two stable states is its mean velocity profile, which is a function of the coordinate characterizing the transition. Indeed, this quantity provides information on local variations in the speed of the transition paths including the effect of stochastic reversals opposite to their average direction due thermal fluctuations and the detailed barrier shape. Remarkably, the mean velocity along transition paths  establishes a link between the mean transition-path time, the probability density of transition-path points, the equilibrium probability density of the system's coordinate, and the equilibrium reactive flux of the transition \cite{hummer2004,berezhkovskii2018}. Therefore, it is of paramount importance to measure in a well controlled experiment the mean transition path velocity in order to verify its theoretically predicted properties as well as its relation with other relevant quantities describing thermally activated processes.

In this paper, we present an experimental study of the transition path dynamics of a colloidal bead hopping between two optical potential wells. Our main goal is to directly measure the mean velocity profile of the transition paths of the particle position in the transition region over the energy barrier, and to probe its relationship with other quantities characterizing the dynamics of the system. To this end, unlike most of the previous investigations using colloidal particles moving in water across bistable potentials, here we perform experiments in a glycerol/water mixture, whose higher viscosity leads to a slower particle motion, thus enabling us to spatially resolve the mean transition-path velocity. In turn, this also allows for the the investigation of the effect of the sampling frequency on the calculation of such mean velocity, which clearly demonstrates that it can be underestimated if the time resolution is not at least two orders of magnitude smaller than the mean transition path time. Moreover, our experimental results confirm that the mean transition path velocity provides a fundamental link between the mean transition path times and the mean escape times of thermally activated processes.


\section{Experimental description}\label{sect:exp}

\begin{figure*}[htb]
 \centering
\includegraphics[width=1.\columnwidth]{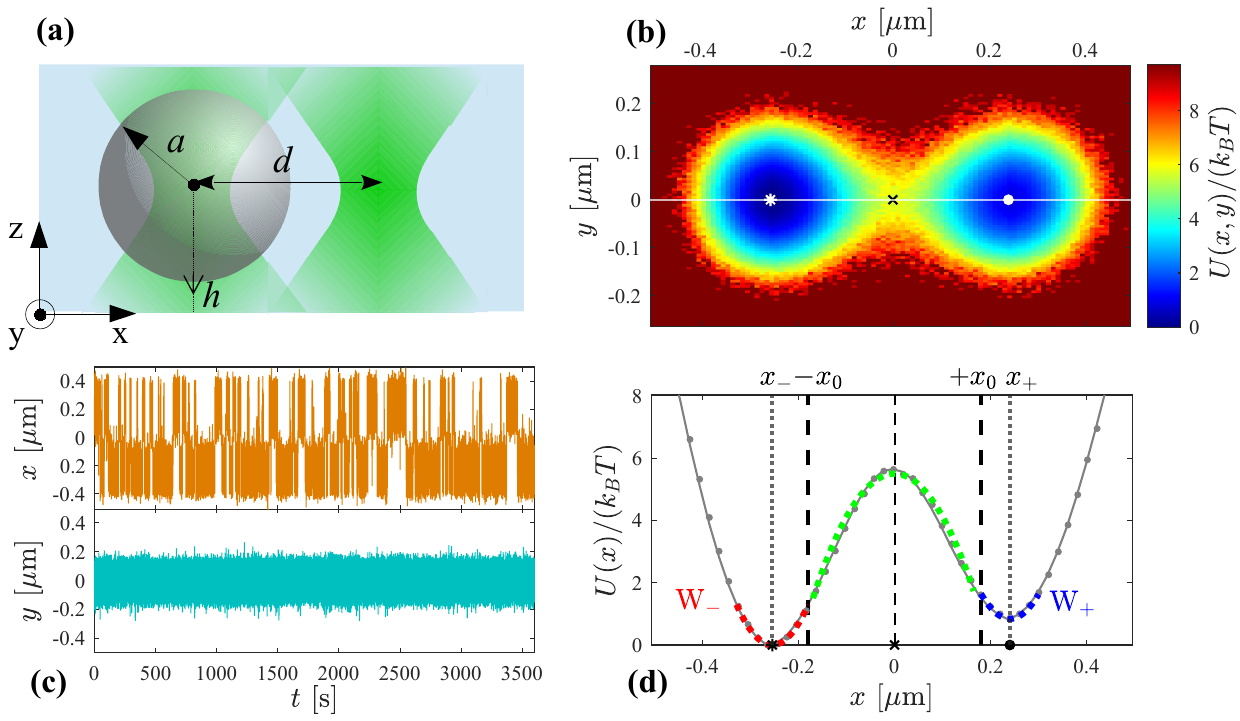}
\caption{(a) Sketch of a spherical bead of radius $a = 0.25\,\mu$m moving in a glycerol/water mixture across a bistable optical potential created by means of two optical tweezers with separation $d \approx 0.5 \, \mu$m, which keep it at a distance $h \approx 10\,\mu$m from the lower cell wall. The coordinate system used in the data analysis is also depicted. 
(b) Color map  of the two-dimensional optical potential $U(x,y)$ acting on the bead motion in the x-y plane perpendicular to the direction of propagation of the laser beams (z). The solid straight line represents the x-axis ($y = 0$) connecting the minimum of the left potential well ($*$), the maximum of the barrier ($\times$), and the minimum of the right well ($\mdsmblkcircle$). (c) Example of the stochastic time evolution over 1 hour of the coordinates of the bead's center of mass, $x(t)$ and $y(t)$, in the plane x-y. (d) Spatial profile of the one-dimensional potential $U(x)$ determined by means of eqn (\ref{eq:1Dpotential}) via the marginal probability density of $x$ (solid line), and by evaluating the two-dimensional potential along the x-axis, $U(x,y=0)$ (dots). The colored dotted lines represent the parabolic fits of $U(x)$ given by eqn (\ref{eq:parabol}) in the regions $[x_{\pm} - \delta x_{\pm}, x_{\pm} + \delta x_{\pm}]$ around the minima at $x = x_{\pm}$ of the two wells, W$_{\pm}$, where $\delta x_{\pm} = |x_{\pm} \mp x_0|$, and the inverted parabolic fit described by eqn (\ref{eq:invparabol}) of the barrier inside the transition region $[-x_0, + x_0]$.}
 \label{fig:1}
\end{figure*}

\subsection{Experimental setup}\label{subsec:setup}

The experiments are conducted using a very dilute dispersion of spherical silica beads (diameter $2a = 0.5\,\mu$m) in a liquid mixture of glycerol and ultrapure water at 52.5\%~wt (less than 1 colloidal particle in 1 nl of solution). The colloidal dispersion is confined in a sample cell composed of a microscope glass slide that is parallelly stuck to a coverslip by double-sided adhesive tape (separation $\sim 100\,\mu$m), and sealed with epoxy glue to avoid leakage and evaporation of the liquid. Two optical tweezers are created inside the sample cell by tightly focusing two orthogonally polarized laser beams (wavelength $\lambda = 532$~nm, total power of 70~mW) by means of an oil-immersion objective ($100 \times$, numerical aperture NA $= 1.3$). Then, as sketched in Fig. \ref{fig:1}(a), a single bead is trapped by the optical tweezers at $h \approx 10\,\mu$m away from the lower wall of the sample cell and very far from any other particle in order to avoid the effect of hydrodynamic interactions on the particle motion. The experiments are carried out at room temperature ($T = 22 \pm 0.1^{\circ}$C) at which the dynamic viscosity of the glycerol/water mixture is $\eta = 0.0061 \pm 0.0017$~Pa~s \footnote[2]{This value was determined in situ from the mean square displacement of the bead in a single optical trap, which was achieved by simply blocking the second laser beam.}, which is six times larger than the viscosity of pure water at the same temperature. Accordingly, the diffusive dynamics of the bead becomes six times slower in the glycerol/water mixture than in water. The separation between the two optical traps is fixed at $d \approx 0.5\,\mu$m, at which we clearly observe that the particle randomly hops between the two minima of the resulting optical potential. We record independent videos of 4 distinct individual particles moving in such a bistable energy landscape using a CMOS camera at a sampling frequency of $f_0 = 1500$ frames per second during 1 hour, over which each particle transits alternately between the two potential wells approximately 300 times. From the recorded videos, we detect the coordinates $(x,y)$ of the center of mass of the two-dimensional projection of the distinct spherical particles perpendicular to the direction of propagation of the laser beams, z, by means of standard particle-tracking routines with a spatial resolution of 5 nm.

\subsection{Potential landscape}\label{subsec:potential}

We reconstruct the detailed shape of the effective optical potential $U(x,y)$ acting on the two-dimensional motion of each particle by use the equilibrium Boltzmann distribution
\begin{equation}\label{eq:Boltzman}
    \rho_{eq}(x,y) = \rho_0 \exp\left[-\frac{U(x,y)}{k_B T} \right],
\end{equation}
where $\rho_0$ is a normalization constant and $\rho_{eq}(x,y)$ is simply computed from the two-dimensional histogram of the tracked coordinates of the particle over one hour. An example of the profile of $U(x,y)$ is illustrated in Fig. \ref{fig:1}(b), which features two neatly defined wells, whose minima are located at $(-0.253\,\mu\mathrm{m},0)$ and $(+0.242\,\mu\mathrm{m},0)$, respectively, separated by an energy barrier of height $\sim 5 k_B T$ with a maximum at $(0,0)$. This potential is symmetric with respect to the straight line connecting the two minima of the wells and the maximum of the barrier, which corresponds to the x-axis $(y = 0)$. An example of the stochastic time evolution of the position of the bead moving in this biestable optical potential is shown in Fig. \ref{fig:1}(c). We observe that, while the coordinate $x(t)$ clearly mirrors the thermally activated transitions of the bead across the bistable potential, the coordinate $y(t)$ does not exhibit any signature of hopping dynamics. Therefore, we can restrict the analysis of the barrier crossing process of the bead to its coordinate $x$. From the marginal probability density $\rho_{eq}(x) = \int_{-\infty}^{\infty} dy \rho_{eq}(x,y)$, which is computed from the  histogram of $x$, we determine the one-dimensional potential
\begin{equation}\label{eq:1Dpotential}
    U(x) = -k_B T \ln \rho_{eq}(x) + u_0,
\end{equation}
where $k_B$ is the Boltzmann constant and $u_0$ is an arbitrary constant. In Fig. \ref{fig:1}(d) we plot the profile of $U(x)$ that is determined from the same data used for the calculation of the two-dimensional potential shown in Fig. \ref{fig:1}(b), which displays two potential wells, W$_-$ and W$_+$, and a barrier, whose corresponding minima and maximum  are located at the same positions as those of $U(x,y)$ along the x-axis, \textit{i.e.} at $x = x_{\pm}$ and $x = 0$, respectively, where $x_- = -0.253 \, \mu$m and $x_+ = +0.242 \, \mu$m. Moreover, as verified in Fig. \ref{fig:1}(d), the complete one dimensional profile of $U(x)$ coincides with that of $U(x,y=0)$ along the x-axis, which demonstrates that $x(t)$ is actually decoupled from $y(t)$, thus being an optimal reaction coordinate to fully describe the activated transitions of the bead.
Furthermore, we notice that the potential wells are in general asymmetric, whose depths can differ from each other up to $\sim 1 k_B T$ depending on the analyzed particle. Therefore, the escape events of the particle over the barrier starting from W$_-$  must be studied separately from those taking place from W$_+$ in order to correctly compute their corresponding rates. Accordingly, as depicted in Fig. \ref{fig:1}(d), for the analysis of the Kramers escape process we perform a parabolic fit around each minimum at $x = x_{\pm}$ of the potential wells 
\begin{equation}\label{eq:parabol}
    U(x) = U(x_{\pm}) + \frac{1}{2} \kappa_{\pm} (x-x_{\pm})^2, 
\end{equation}
where $\kappa_{\pm} > 0$ are the corresponding local curvatures of the potential. On the other hand, we define the boundaries of the transition region at $x = -x_0$ and $x = +x_0$, \textit{i.e.} symmetrically positioned with respect to barrier maximum at $x = 0$, by choosing the largest value of $x_0$ such that the potential $U(x)$ can be approximated as an inverted parabola within $-x_0 \le x \le +x_0$
\begin{equation}\label{eq:invparabol}
    U(x) = U(0) - \frac{1}{2}\kappa_0 x^2, 
\end{equation}
with $\kappa_0 > 0$, as depicted in Fig. \ref{fig:1}(d). We check that the value $x_0 = 0.18\,\mu$m allows for the above-mentioned parabolic fit of the energy barrier for all the particles used in the experiments even though the values of $\kappa_0$ slightly vary between distinct particles. It should also be noted that the symmetry of $U(x)$ with respect to $x = 0$ for $-x_0 \le x \le +x_0$  permits to study the duration of transition paths just by paying attention to the direction of the transition regardless of the initial well from which the particle enters the transition region.

\subsection{Characteristic time-scales}\label{subsec:times}

\begin{figure*}[ht]
\centering
\includegraphics[width=1.\columnwidth]{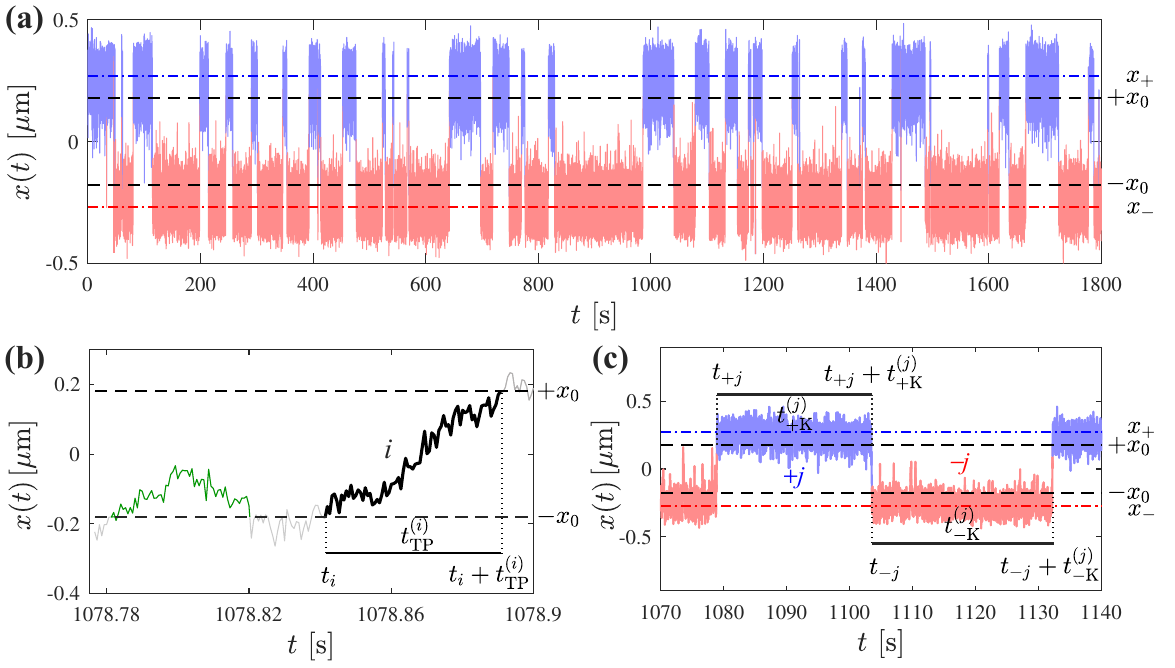}
  \caption{(a) Example of the time evolution of the coordinate $x(t)$ reflecting the stochastic hopping of the bead during 30 minutes between the potential wells W$_-$ and W$_+$ with minima at $x_-$ and $x_+$. Red and blue segments of the trajectory represent events where the bead resides in wells W$_-$ and W$_+$, respectively, until it fully escapes by thermal activation to the contiguous well by following a transition path through the transition region $[-x_0, +x_0]$. (b) Diagram of the calculation of the experimental values taken along the bead trajectory by the transition path time, $t_{\mathrm{TP}}$, defined inside the transition region $[-x_0, +x_0]$. The solid thick black line represents an actual transition path, whereas the solid thin green line corresponds to a failed attempt that does not count as such.  (c) Diagram of the calculation of the experimental values taken along the bead trajectory by the escape times, $t_{\mathrm{+K}}$ and $t_{\mathrm{-K}}$, from the potential wells W$_+$ and W$_-$, which are determined from the duration of the segments marked in blue and red, respectively. In Figs. \ref{fig:2}(a)-\ref{fig:2}(c) the dashed lines outline the boundaries of the transition region, and the dotted-dashed lines mark the location of the minima of the potential wells.}
  \label{fig:2}
\end{figure*}

Fig. \ref{fig:2} details the calculation of the two main time-scales characterizing the barrier-crossing process of the bead, namely, the mean escape time, whose inverse is equal to the corresponding transition rate, and the mean transition path time. First, in Fig. \ref{fig:2}(a) we depict an expanded view over 30 minutes of the hopping dynamics of the bead through its coordinate $x(t)$, which displays comparatively long stays of random duration represented in different colors, corresponding to situations where the particle equilibrates in each potential well. These long-lived events are interrupted by much shorter excursions of the bead deep inside the transition region, most of then failing to reach the boundary on the other side of the barrier, and some of them corresponding to actual transition paths along which the bead fully traverses from one potential well to the other. To measure the mean transition path time, we determine the duration $t_{\mathrm{TP}}^{(i)}$ of the $i-$th segment of the particle trajectory that enters the transition region at $x = \pm x_0$ and exits at $x = \mp x_0$ without crossing back into the entrance point, \textit{i.e.} by definition a transition path, as exemplified by the portion $\{ x(t); t_i \le t \le t_i + t_{\mathrm{TP}}^{(i)} \}$ represented as a solid thick black line in Fig. \ref{fig:2}(b). This provides the particular value $t_{\mathrm{TP}}^{(i)}$ taken by the transition path time, $t_{\mathrm{TP}}$, which is a stochastic variable due to the effect of thermal fluctuations on the particle motion. This definition excludes events starting at $x = \pm x_0$ and returning to the same point before reaching the opposite boundary at $x = \mp x_0$, like the part of the trajectory depicted as a solid thin green line in Fig. \ref{fig:2}(b). Once all the transition paths performed by the bead within the transition region have been identified and the their corresponding duration has been determined, we calculate their arithmetic mean 
\begin{equation}\label{eq:meantTP}
\langle t_{\mathrm{TP}} \rangle = \frac{1}{N_0}\sum_{i=1}^{N_0} t_{\mathrm{TP}}^{(i)},
\end{equation}
where $N_0$ is the total number of transitions paths detected along the complete trajectory over 1 hour, typically $N_0 \sim 300$. On the other hand, to
compute the mean escape times, we must find the particular values taken during the experiment by the escape times $t_{+ \mathrm{K}}$ and $t_{- \mathrm{K}}$, which are also stochastic variables, where the signs $+$ and $-$ represent the well from which the particle escapes, \textit{i.e.} W$_+$ or W$_-$, respectively, whereas K stands for ``Kramers' escape process''. For the sake of brevity, in the following we will use the notation $\pm$ to simultaneously refer to both escape processes and to the signs associated to the parameters characterizing the particle escape from the corresponding well. For this calculation, we start counting the cumulative time spent by the particle around the respective minimum at $x = x_{\pm}$ since it left the transition region at time $t=t_{\pm j}$ from $x = \pm x_0$, including all failed attempts to reach the point $x = \mp x_0$ on the other side of the barrier, as illustrated in Fig. \ref{fig:2}(c) by the $\pm j-$th segment $\{x(t); t_{\pm j} \le t \le t_{\pm j} + t_{\pm \mathrm{K}}^{(j)} \}$ shown in blue ($+$) and red ($-$). Once the particle performs a transition path by successfully traversing from $x = \pm x_0$ to $x = \mp x_0$, we consider that the $\pm j-$th escape event has been completed and then we stop counting the time elapsed since $t_{\pm j}$, thus yielding the corresponding value $t_{\pm \mathrm{K}}^{(j)}$ of $t_{\pm\mathrm{K}}$. By computing the values of $t_{\mathrm{\pm K}}$ for all the $N_{\pm}$ escape events occurring along the experimental trajectory $x(t)$, typically $N_+ \sim 150 \sim N_-$, we simply calculate their arithmetic average, thus leading to the values of the mean escape times from the corresponding wells W$_{\pm}$, \textit{i.e.} 
\begin{equation}\label{eq:meantK}
\langle t_{\pm \mathrm{K}} \rangle = \frac{1}{N_{\pm}} \sum_{i=1}^{N_{\pm}} t_{\pm \mathrm{K}}^{(j)}.
\end{equation}
It should be pointed out that, according to the previous definition, the mean escape time of the bead from the well W$_{\pm}$ is equal to its mean first passage time from $x = \pm x_0$ to $x = \mp x_0$ under the condition that the particle initially exited the transition region through $x = \pm x_0$ rather than entering it.

\section{Model}\label{sect:model}

We now consider a stochastic model for the hopping dynamics of the bead along the x-axis across the effective bistable potential $U(x)$. Since the particle moves in a highly viscous medium at constant temperature, the  probability density $\rho(x,t|x',t')$ to locate it at position $x$ at time $t$ given that it was at $x'$ at time $t' < t$ is governed by the Smoluchowski equation
\begin{equation}\label{eq:Smoleq}
\frac{\partial \rho(x,t|x',t')} {\partial t} =  D \frac{\partial}{\partial x} \left\{ e^{ -\frac{U(x)}{k_B T} } \frac{\partial}{\partial x} \left[ e^{\frac{U(x)}{k_B T}} \rho(x,t|x',t') \right] \right\}.
\end{equation}
In eqn (\ref{eq:Smoleq}), $D = k_B T / \gamma$ is the diffusion coefficient of the particle in the viscous medium, where $\gamma = 6\pi a \eta$ is its friction coefficient. From the probability flux associated to eqn (\ref{eq:Smoleq}), \textit{i.e.} $J(x',t) = -D e^{-{U(x)}/(k_B T)} \partial_x \left[e^{{U(x)}/(k_B T)} \rho(x,t|x',t' = 0)\right]|_{x = +x_0}$, originating from a point $ x' > -x_0$ within the transition region and passing through the boundary $x = +x_0$, the well-known analytic approximation for the mean transition path time can be found by taking the limit $x' \rightarrow -x_0$
\cite{hummer2004,chung2014}
\begin{equation}\label{eq:meanTPth}
    \langle t_{\mathrm{TP}} \rangle = \tau_0 \ln \left( 2e^{\gamma_{\mathrm{EM}}} \frac{\delta U}{k_B T} \right).
\end{equation}
In eqn (\ref{eq:meanTPth}), $\delta U = \frac{1}{2} \kappa_0 x_0^2$ is the height of the barrier with respect to the boundaries of the transition region, $\gamma_{\mathrm{EM}} \approx 0.5772$ is the Euler–Mascheroni constant, and $\tau_0 = \gamma / \kappa_0$ is the characteristic time of the particle when moving in the unstable region on top of the barrier. This approximation for $\langle t_{\mathrm{TP}} \rangle$ is valid provided that the relative barrier height $\delta U$ is sufficiently large in comparison with $k_B T$. Likewise, from the steady-state solution of eqn (\ref{eq:Smoleq}), the parabolic approximations of the potential around the minima at $x = x_{\pm}$ and the top of the barrier at $x = 0$ described by eqns (\ref{eq:parabol}) and (\ref{eq:invparabol}), respectively, lead to Kramers' formula for the mean escape times from each potential well \cite{kramers1940}
\begin{equation}\label{eq:meantKth}
    \langle t_{\pm \mathrm{K}} \rangle = 2\pi \sqrt{\tau_0 \tau_{\pm}} \exp \left( \frac{\Delta U_{\pm}}{k_B T} \right).
\end{equation}
In eqn (\ref{eq:meantKth}), $\tau_{\pm} = \gamma / \kappa_{\pm}$ are the viscous relaxation times of the particle in the harmonic regions around the potential minima located at $x = x_{\pm}$, whereas $\Delta U_{\pm}$ denotes the heights of the energy barrier relative to the minima of the wells W$_{\pm}$, \textit{i.e.} $\Delta U_{\pm} = U(0) - U(x_{\pm})$. Eqn (\ref{eq:meantKth}) is valid if the barrier is sufficiently high with respect to $k_B T$, which guarantees that once the particle arrives at a certain well after completing a transition path, it has enough time to equilibrate there before crossing back over the barrier.

Apart from the previous characteristic time-scales, a quantity that provides more detailed statistical information on the spatial variations of the speed of an activated transition is the local mean velocity along transition paths. By counting the total time spent on average by a single transition path within the spatial range $[x -\Delta x/2 , x + \Delta x/2 ]$ fully contained in the transition region, $\langle \Delta t(x,\Delta x) \rangle$, including the duration of its recrossing into this interval due to the stochastic reversals opposing the average direction of the transition, its mean velocity at $x$ is defined as
\begin{equation}\label{eq:tpvel}
    v_{\mathrm{TP}}(x) = \lim_{\Delta x \rightarrow 0} \frac{\Delta x}{\langle \Delta t(x,\Delta x) \rangle}.
\end{equation}
Moreover, from the definition of $v_{\mathrm{TP}}(x)$ it should be noted that the time that a transition path needs on average to travel from $x$ to $x+dx$ is $dx/v_{\mathrm{TP}}(x)$, and then the fraction of time spent there with respect to its total average duration, $\langle t_{\mathrm{TP}} \rangle$, is $dx/\left[v_{\mathrm{TP}}(x) \langle t_{\mathrm{TP}} \rangle \right]$. This must be equal to the probability that the particles follows a transition path inside that infinitesimal region during the same time interval, which is simply given by $\rho(x|\mathrm{TP})dx$, where $\rho(x|\mathrm{TP})$ is the probability density function of the position $x$ within the transition region conditioned to belong only to a transition path. Therefore, the mean transition-path velocity must satisfy the following general relationship with the probability density of the transition path positions across the entire transition region
\begin{equation}\label{eq:meanvel0}
   \rho (x|\mathrm{TP}) = \frac{1}{\langle t_{\mathrm{TP}} \rangle v_{\mathrm{TP}}(x)}.
\end{equation}
As demonstrated in \cite{hummer2004}, for the diffusive dynamics over the barrier of $U(x)$ described by eqn (\ref{eq:Smoleq}), one can find general expressions for $\langle t_{\mathrm{TP}} \rangle$ and $\rho (x|\mathrm{TP})$ in terms of the splitting probabilities $\Phi_{\pm x_0}(x)$, defined as the probabilities that a path starting at $x$ will reach the boundaries $\pm x_0$, respectively. By definition, such splitting probabilities satisfy the condition $\Phi_{-x_0}(x) + \Phi_{+x_0}(x) = 1 $, and can be fully expressed in terms of the equilibrium Boltzmann distribution within the transition region $-x_0 < x' < +x_0$, $\rho_{eq}(x') \propto \exp\left[-U(x')/(k_B T)\right]$, as
\begin{equation}\label{eq:split}
    \Phi_{+x_0}(x) = \frac{\int_{-x_0}^x dx' \left[ \rho_{eq}(x')\right]^{-1}}{\int_{-x_0}^{+x_0} dx' \left[ \rho_{eq}(x')\right]^{-1}}.
\end{equation}
Then, by use of eqn (\ref{eq:meanvel0}), the local mean velocity of transition paths can be recast as 
\begin{equation}\label{eq:meanvelinv}
   v_{\mathrm{TP}}(x) = D \left[ \int_{-x_0}^{x_0}\frac{dx'}{ \rho_{eq}(x')} \Phi_{+x_0}(x)\Phi_{-x_0}(x)\rho_{eq}(x) \right]^{-1}, 
\end{equation}
which allows one to find the following analytic expression for its spatial profile in the case of the inverted parabolic approximation of the barrier inside the transition region, $U(x) = U(0) - \frac{1}{2}\kappa_0 x^2$
\begin{equation}\label{eq:meanvel}
    v_{\mathrm{TP}}(x) =  \sqrt{\frac{8 D}{\pi \tau_0}} \frac{\mathrm{erf} \left( \sqrt{\frac{\delta U}{k_B T}}\right)\exp\left( -\frac{\kappa_0 x^2}{2k_B T}\right)}{\mathrm{erf}^2\left( \sqrt{\frac{\delta U}{k_B T}}\right) - \mathrm{erf}^2 \left( \sqrt{\frac{\kappa_0}{2k_B T}} x \right)}.
\end{equation}

Furthermore, as discussed in \cite{hummer2004,berezhkovskii2018}, a remarkable consequence of the properties of the local mean velocity, or equivalently the conditional probability density of the transition path position, is that it establishes direct link between the mean transition-path times and the transition rates through equilibrium and transition path distributions. Indeed, by use of Bayes' theorem, the probability density function of the particle position along a transition path going from $-x_0$ to $+x_0$ can be expressed as
\begin{equation}\label{eq:Bayes}
    \rho (x|\mathrm{TP}) = \frac{P(\mathrm{TP}|x) \rho_{eq} (x)}{P(\mathrm{TP})},
\end{equation}
where $P(\mathrm{TP}|x)$ is the probability that at position $x$ in the transition region the particle is on a transition path from $-x_0$ to $+x_0$, and $P(\mathrm{TP})$ is the probability that it has undergone through a transition path. Since the latter represents the fraction of the total time spent by the particle in a transition path from $-x_0$ to $+x_0$, it can be expressed as $P(\mathrm{TP}) = J_{-} \langle t_{\mathrm{TP}}\rangle $, where $J_{-} = P_{-}/\langle t_{-\mathrm{K}}\rangle$ is the steady-state flux of the transition and $P_-$ is the probability of finding the particle in the well W$_-$, \textit{i.e.} at $x < -x_0$. Similar relations hold for the transition paths from $+x_0$ to $-x_0$, thus yielding the general expression for the ratio of the characteristic time-scales of the barrier crossing process
\begin{equation}\label{eq:ratiotime}
    \frac{\langle t_{\mathrm{TP}}\rangle}{\langle t_{\pm \mathrm{K}} \rangle} = \frac{P(\mathrm{TP}|x) \rho_{eq} (x)}{P_{\pm} \rho (x|\mathrm{TP})},
\end{equation}
which is independent of $x$ even though the functions on the right-hand side depend separately on $x$ in a non-trivial fashion.

\section{Results}\label{sect:res}

\begin{figure*}[ht]
 \centering \includegraphics[width=0.9\columnwidth]{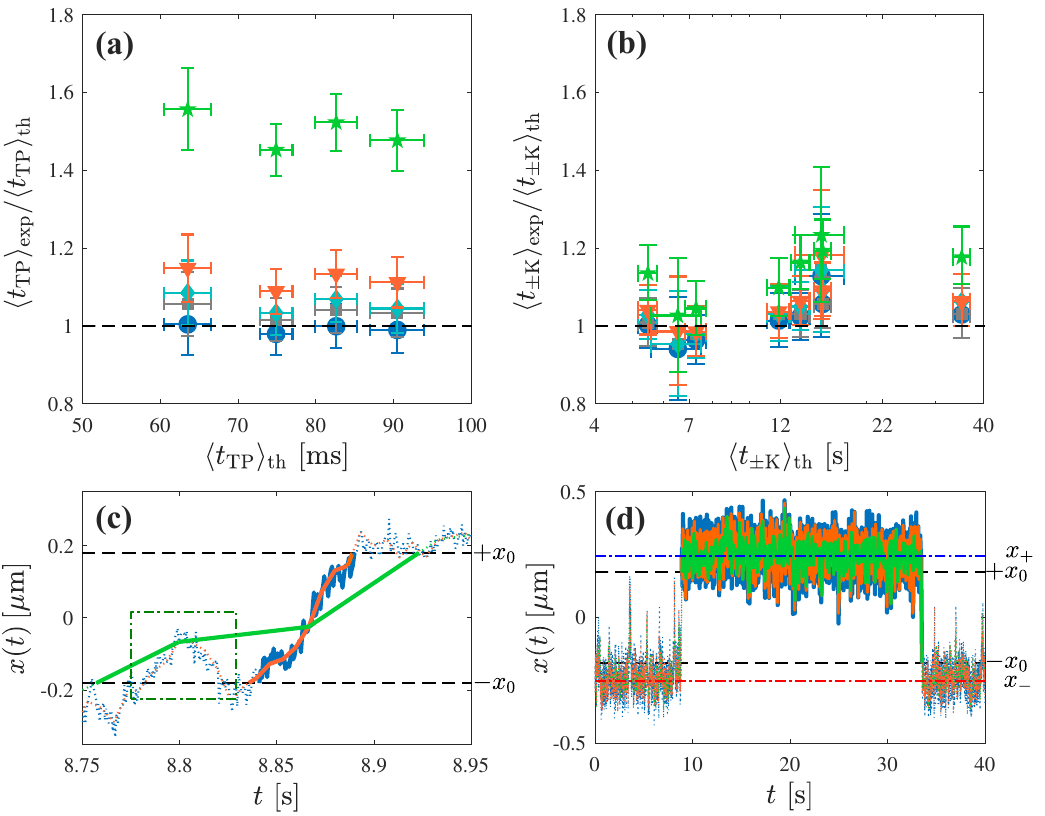}
\caption{(a) Ratio of the experimentally measured values of the mean transition path time for distinct colloidal beads using eqn (\ref{eq:meantTP}), $\langle t_{\mathrm{TP}}\rangle_{\mathrm{exp}}$, to the corresponding theoretical prediction, $\langle t_{\mathrm{TP}}\rangle_{\mathrm{th}}$, given by eqn (\ref{eq:meanTPth}). (b) Ratio of the experimentally measured values of the mean escape time for distinct colloidal beads using eqn (\ref{eq:meantK}), $\langle t_{\mathrm{\pm K}}\rangle_{\mathrm{exp}}$, to the corresponding theoretical prediction, $\langle t_{\mathrm{\pm K}}\rangle_{\mathrm{th}}$, according to eqn (\ref{eq:meantKth}). The distinct symbols in Figs. \ref{fig:3}(a) and \ref{fig:3}(b) represent the results of decimating the original data that was experimentally recorded at a sampling frequency $f_0 = 1500$~Hz ($\bigcirc$) to a reduced frequency $f_0 / n$, with $n=2$ ($\smwhtsquare$), $n = 5$ ($\diamondsuit$), $n = 10$ ($\triangledown$), and $n = 100$ ($\bigwhitestar$). (c) Example of a transition path identified along the original trajectory that was recorded at $f_0 = 1500$~Hz (solid blue line) and those apparently detected when decimating the original data to $f_0/n$, with $n =10$ (solid orange line) and $n =100$ (solid green line). The dotted lines represent the parts of the trajectory that are not considered as transition paths at the corresponding sampling frequency, which are plotted using the same color code as that of the transition paths. The dotted-dashed line rectangle encloses an unsuccessful attempt of the particle inside the transition region to cross the barrier, which is misidentified as part of a transition path when sampled at $f_0/100$. (d) Example of a segment of the trajectory corresponding to a full residence and  escape event of the particle from the well W$_+$ (solid lines) and those that do not belong to it (dotted lines). Same color code for the sampling frequency as that in Fig. \ref{fig:3}(c). In Figs. \ref{fig:3}(c)-\ref{fig:3}(d) the dashed lines outline the boundaries of the transition region, and the dotted-dashed lines mark the location of the minima of the potential wells.}
 \label{fig:3}
\end{figure*}

In this Section, using the experimental trajectories of the bead, we test the validity of the general expressions given in eqns (\ref{eq:meanvel0}) and (\ref{eq:ratiotime}), both involving the determination of the spatial profile of its mean transition path velocity. To this end, using eqns (\ref{eq:meantTP}) and (\ref{eq:meantK}) we first compute the mean transition path time of the bead and its two mean escape times across the bistable potential landscape at the experimental sampling frequency $f _0 = 1500$~Hz, and then examine how these quantities are affected upon decimation of the original data. In Figs. \ref{fig:3}(a) and \ref{fig:3}(b) we show that at $f_0$, different experiments corresponding to distinct beads moving in their respective optical potentials exhibit mean transition path times and mean escape times from each well that are in quantitative agreement with eqns (\ref{eq:meanTPth}) and (\ref{eq:meantKth}), respectively. For the theoretical calculations by means of these equations, we use the values of the parameters $\kappa_0$, $\kappa_{\pm}$, $\delta U$, and $\Delta U_{\pm}$ that were determined from the analysis described in Subsection \ref{subsec:potential} of the coordinate $x(t)$ recorded at the original frequency $f_0$. We point out that the temporal resolution given by $1/f_0 = 0.67$~ms, at which there is an excellent agreement between the measured values of $\langle t_{\mathrm{TP}}\rangle$ and $\langle t_{\pm \mathrm{K}}\rangle$ and those theoretically predicted by eqns (\ref{eq:meanTPth}) and (\ref{eq:meantKth}), is two order of magnitude shorter than the predicted values of the mean transition path time ($60\,\mathrm{ms} < \langle t_{\mathrm{TP}}\rangle < 100\,\mathrm{ms}$), see Fig. \ref{fig:3}(a). Reducing the sampling frequency of the trajectories by less than one order of magnitude upon decimation of the original data by a factor $n < 10$, \textit{i.e.} $f_0 / n > f_0/10$, gives rise to a slight increase in $\langle t_{\mathrm{TP}}\rangle$ of at most ten percent, as can be seen in Fig. \ref{fig:3}(a) for $n = 2$ and $n = 5$. Moreover, a further increase in the decimation factor $n$ by more than one order of magnitude, $f_0 / n < f_0/10$, leads to systematic deviations of the measured values of $\langle t_{\mathrm{TP}}\rangle$ from the theoretical one, as shown in Fig. \ref{fig:3}(a) for $n = 10$ and $n = 100$. In such cases, it can be seen that the resulting values upon decimation are up to 60~\% larger than those measured at $f_0$. On the contrary, the mean escape time is less affected by the decimation of the original trajectory, as shown in Fig. \ref{fig:3}(b), where an increase of at most 20~\% of the experimental value with respect to the theoretical estimate occurs when reducing the sampling frequency by $n = 100$. The main reason of the significant increase in the measured values of $\langle t_{\mathrm{TP}}\rangle$ when reducing the sampling frequency by more than one order of magnitude is illustrated in Fig. \ref{fig:3}(c), where we trace a transition path detected at the original frequency $f_0$, as well as the same path apparently recognized after decimating the original data by a factor $n = 10$ and $n = 100$. In particular, for $n = 100$ it becomes clear that unsuccessful attempts of the particle to surmount the barrier occurring inside the transition region can be incorrectly identified as part of a transition path due to an excessive decimation of the data points. It should also be noted that in this case the time resolution ($100/f_0 = 67$~ms) is of the same order of magnitude of the actual values of $\langle t_{\mathrm{TP}}\rangle$ either calculated via eqn (\ref{eq:meanTPth}) or measured at $f_0$, thus resulting in a number of data points belonging to a single transition path of the order of one. On the other hand, since the measurement of $\langle t_{\pm \mathrm{K}}\rangle$ from the experimental trajectories involves the detection of the residence of the particle in a given well and the subsequent transition path through which it escapes, it is less prone to artifacts when analyzed at reduced sampling rate. This is because the overestimate of the duration of the transition path, which occurs mostly by misidentifying paths recrossing the boundaries of the transition region as previously discussed, represents only a small fraction of the time that the particle previously stayed in the well. Indeed, this is exemplified by the dwelling and later escape of the particle from the well W$_+$ depicted in Fig. \ref{fig:3}(d), where it can be observed that the residence time in the well is barely affected by the decimation even for $n=100$, whereas the duration of the transition path that triggers the escape is much shorter than the total time that the particle needs to wait to fully traverse the barrier. Hence, $\langle t_{\pm \mathrm{K}}\rangle$ can still be well estimated in practice even at reduced sampling frequencies provided the resulting time resolution is sufficient for the detection of transition paths, \textit{i.e.} at least of the order of $\langle t_{\mathrm{TP}}\rangle$. Therefore, the results of this analysis reveal that artifacts can easily arise mainly in the measured values of $\langle t_{\mathrm{TP}}\rangle$ if the number of detected data points per transition path is small, which must also translate into unphysical artifacts in the reconstructed shape of the mean transition path velocities, as shown below. 

\begin{figure*}[htb]
 \centering
\includegraphics[width=1.\columnwidth]{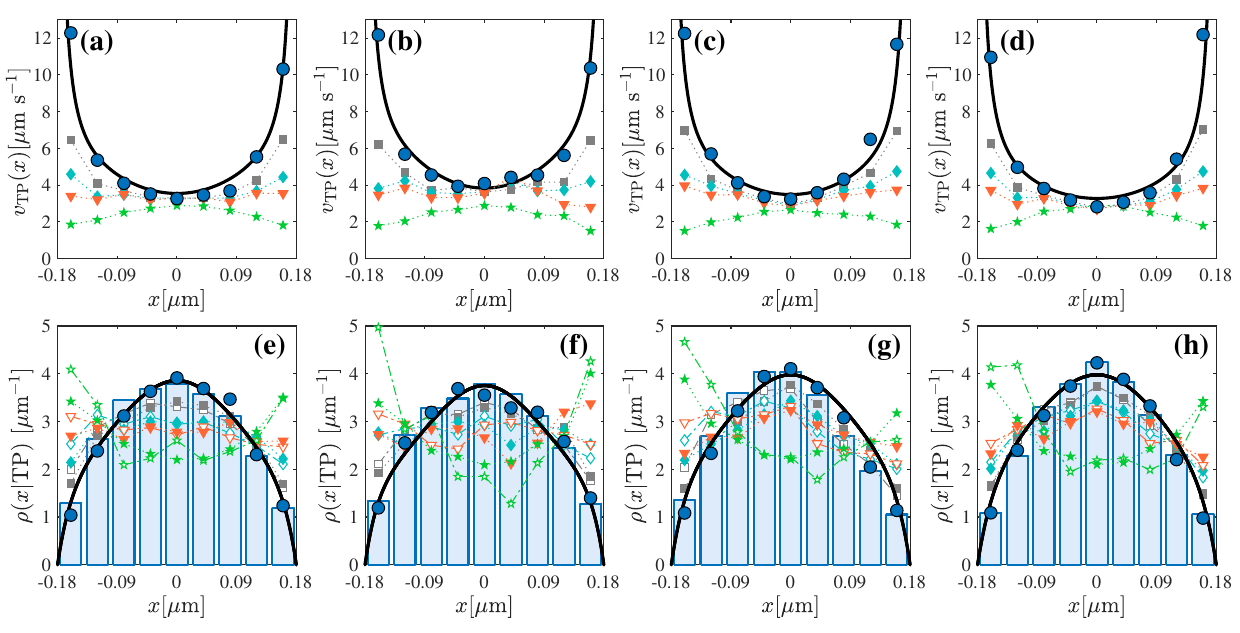}
\caption{(a)-(d) Mean transition path velocities of the four analyzed particles, $v_{\mathrm{TP}}(x)$, moving in their respective optical potentials, where the shape of their energy barriers can be approximated as inverted parabolas with curvatures (a) $\kappa_0=9.57 \pm 0.397 \times 10^{-7} {\mathrm{N}}\, \mathrm{m}^{-1}$, (b) $\kappa_0=1.127\pm 0.0460\times 10^{-6} {\mathrm{N}}\, \mathrm{m}^{-1}$, (c) $\kappa_0=9.247\pm 0.282\times 10^{-7} {\mathrm{N}}\, \mathrm{m}^{-1}$, and (d) $\kappa_0=8.08 \pm 0.271 \times 10^{-7} {\mathrm{N}}\, \mathrm{m}^{-1}$. The different symbols represent the results from the experimental data recorded at a sampling frequency $f_0 = 1500$~Hz ($\bigcirc$) and after decimation by a factor $n=2$ ($\smwhtsquare$), $n = 5$ ($\diamondsuit$), $n = 10$ ($\triangledown$), and $n = 100$ ($\bigwhitestar$). The thick solid line depicts the analytic expression of $v_{\mathrm{TP}}(x)$ given by eqn (\ref{eq:meanvel}), whereas the dotted lines are just guides to the eye. (e)-(h) Probability density functions of the transition path position of the four analyzed particles, $\rho (x|\mathrm{TP})$, which are directly computed from the histogram of experimental data points belonging only to transition paths (vertical bars), and indirectly by means of eqn (\ref{eq:meanvel0}) using the measured values of mean transition path times and the mean velocities profiles plotted in Figs \ref{fig:3}(a) and \ref{fig:4}(a)-\ref{fig:4}(d), respectively ($\bigcirc$), both determined from the experimental data recorded at $f_0 = 1500$~Hz. The smaller symbols represent the results after decimating the original data by a factor $n=2$ ($\smwhtsquare$), $n = 5$ ($\diamondsuit$), $n = 10$ ($\triangledown$), and $n = 100$ ($\bigwhitestar$), where the empty symbols correspond to the direct calculation of the histograms of positions along transition paths, and the filled symbols depict the results of the indirect calculation using the corresponding  mean velocity profiles and mean transition path times via eqn (\ref{eq:meanvel0}). The thick solid line represents the analytic expression of $\rho(x|{\mathrm{TP}})$ derived from eqn (\ref{eq:meanvel0}) using eqns (\ref{eq:meanTPth}) and (\ref{eq:meanvel}), whereas the dotted and dotted-dashed lines are just guides to the eye.}
 \label{fig:4}
\end{figure*}

In Figs. \ref{fig:4}(a)- \ref{fig:4}(d) we plot as circles the mean velocity profiles of the transition paths of the four distinct particles analyzed in the transition region over their respective barriers, which are determined from their experimental trajectories recorded at the sampling frequency $f_0 = 1500$~Hz. To calculate $v_{\mathrm{TP}}(x)$ from the particle trajectories by means of eqn. (\ref{eq:tpvel}), we use $\Delta x = 40$~nm, which is selected to be of one order of magnitude larger than the spatial resolution of the detection of $x$ (5~nm) and at the same time one order of magnitude smaller than the total length of the transition region ($2x_0 = 0.36\,\mu$m). In Figs. \ref{fig:4}(a)- \ref{fig:4}(d) we also trace as solid lines the theoretical expression of $v_{\mathrm{TP}}(x)$ derived from the diffusive model of the particle motion that is shown in eqn. (\ref{eq:meanvel}), where the diffusion coefficient $D$ is calculated using the measured value of the viscosity of the glycerol/water mixture, whereas $\kappa_0$ is determined for each specific optical potential, which allows us to compute the corresponding values of $\tau_0$. In all cases, we find an excellent agreement between the experimental mean velocity profiles and their corresponding theoretical expressions, thereby verifying that at a sampling frequency $f_0$ that is at least two orders of magnitude larger than $\langle t_{\mathrm{TP}} \rangle^{-1}$, the spatial dependence of $v_{\mathrm{TP}}(x)$ on the coordinate $x$ can be properly resolved. It should be noted that the shape of $v_{\mathrm{TP}}(x)$ is intuitively consistent with the expected particle dynamics over a parabolic barrier, where a transition path must enter the transition region at a very high speed on average, then becoming slower and slower until it reaches the top of the barrier at $x = 0$, where it attains its smallest value $v_{\mathrm{TP}}(0) \le v_{\mathrm{TP}}(x)$, then becoming increasingly faster when getting away from the barrier maximum to finally leave the transition region. However, if the transitions paths are undersampled at a frequency below $f_0$, artifacts in the mean velocity profile can be observed even at $f_0 / 2$, at which $v_{\mathrm{TP}}(x)$ is underestimated even when it preserves its convex shape across the transition region with a minimum at $x = 0$, as displayed by the squares in Figs \ref{fig:4}(a)- \ref{fig:4}(d). Reducing the experimental sampling frequency leads to a further apparent deformation of the mean velocity profile, which becomes much flatter at $f_0/5$, see the diamonds in Figs. \ref{fig:4}(a)- \ref{fig:4}(d). Note that for these decreased sampling frequencies, $\langle t_{\mathrm{TP}}\rangle$ and $\langle t_{\pm \mathrm{K}}\rangle$ are still in good agreement with their theoretical predictions and with the values determined at $f_0$, thus demonstrating that $v_{\mathrm{TP}}(x)$ is more sensitive to artifacts that prevent a correct reconstruction of its spatial profile. More distorted profiles of the mean transition path velocity can be inaccurately retrieved at smaller sampling frequencies $f_0/n$, as shown in Figs. \ref{fig:4}(a)- \ref{fig:4}(d) for $n = 10$ as triangle symbols, at which $v_{\mathrm{TP}}(x)$ looks uniform across the transition region, and even concave for $n = 100$, as depicted by the star symbols. We point out that the latter represents a totally unphysical behavior of the transition path dynamics over a parabolic barrier, since this represents a situation where the particle would speed up when climbing to the top of the barrier and slowing down past the maximum. This artifact originates mainly from the lower time resolution at $f_0/n$ with $n >1$, which gives rise to an apparent rise in the total measured time $\langle \Delta t(x,\Delta x) \rangle$ that a transition path spends on average in the region $[x-\Delta x/2, x + \Delta/2]$, thus yielding an underestimate of its corresponding mean velocity according to eqn. (\ref{eq:tpvel}). Indeed, with decreasing sampling frequency of the particle motion, longer recrossing events of the neighboring boundary like the section of the trajectory enclosed in the dotted-dashed rectangle in Fig.\ref{fig:3}(c) are falsely detected as part of a nearby transition path, thus seemingly increasing its duration. Note that this artificial increase in $\langle \Delta t(x,\Delta x) \rangle$ at lower temporal resolution becomes more significant close to the boundaries of the transition region than in the barrier maximum, which explains why the reduction in the apparent values of $v_{\mathrm{TP}}(x)$ are smaller around $x = 0$. In turn, this leads to a change of the mean velocity profile from the actual convex profile described by eqn (\ref{eq:meanvel}) when transition paths are correctly identified with negligible recrossing segments, to a concave shape with decreasing sampling rate of the particle trajectories. Therefore, our results highlight again the need of a sampling frequency of at least two orders of magnitude larger than the inverse of the mean transition path time in order to correctly reconstruct the statistical properties of the particle dynamics across the energy barrier. 

In Figs. \ref{fig:4}(e)- \ref{fig:4}(h) we experimentally probe eqn. (\ref{eq:meanvel0}) linking the spatial behavior of the mean transition path velocity, $v_{\mathrm{TP}}(x)$, with the probability density of the transition path position, $\rho (x|\mathrm{TP})$, with a prefactor given by the mean transition path time, $\langle t_{\mathrm{TP}}\rangle$. First, we experimentally determine $\rho (x|\mathrm{TP})$ by computing the normalized histograms of the bead position only along transition paths, which are represented as vertical bars in Figs. \ref{fig:4}(e)-\ref{fig:4}(h). We also employ the measured values of the mean transition path time and the mean velocity profiles that are plotted in Figs \ref{fig:3}(a) and \ref{fig:4}(a)-\ref{fig:4}(d), respectively, in order to determine the right-hand side of eqn (\ref{eq:meanvel0}), which is represented as circles in Figs. \ref{fig:4}(e)- \ref{fig:4}(h). We find that at $f_0 = 1500$~Hz, the relationship shown in eqn (\ref{eq:meanvel0}) holds across the entire transition region, thus verifying its validity in all experiments with different beads. We also check that, at this sampling frequency, the experimental results are quantitatively described by the analytic expression of $\rho (x|\mathrm{TP})$ calculated through eqn (\ref{eq:meanvel0}) using the expressions of $\langle t_{\mathrm{TP}}\rangle$ and $v_{\mathrm{TP}}(x)$ given in eqns (\ref{eq:meanTPth}) and (\ref{eq:meanvel}), respectively, which were derived from the diffusive model of eqn (\ref{eq:Smoleq}) for the particle motion over a parabolic barrier. Nevertheless, we find that artifacts in the shape of $\rho (x|\mathrm{TP})$ arise when reducing the data acquisition frequency to values that are less than two orders of magnitude larger than $\langle t_{\mathrm{TP}}\rangle^{-1}$. This is exemplified by the empty symbols in Figs. \ref{fig:4}(e)-\ref{fig:4}(h) representing the results of the experimental data whose frequency was decimated to $f_0/n$ with $n = 2,5,10,100$, which demonstrates that the profile of $\rho (x|\mathrm{TP})$ obtained by the direct calculation of the histograms of transition path positions is gradually deformed  with increasing $n$, becoming less and less concave for $n = 2$ and $n = 5$, then rather flat for $n = 10$ and convex for $n = 100$. Once again, this artifact stems mainly from the sections of the particle trajectory inside the transition region that are misidentified as part of a transition path at low temporal resolution even when they are not, \textit{e.g.} see the enclosed trajectory segment recorded at $f_0 = 1500$~Hz in Fig. \ref{fig:3}(c), which is incorrectly detected as part of a transition path at $f_0/100 = 15$~Hz. Such events increase the density of data points mainly close to the boundaries $x = \pm x_0$ of the transition region when calculating the histogram of positions along transition paths, thereby leading to an overestimate of $\rho (x|\mathrm{TP})$ in that region and an ensuing underestimate around the top of the barrier. In addition, it must be noticed that when using the experimental values of the mean transition path times displayed in Fig. \ref{fig:3}(a) and the experimental mean velocity profiles plotted in Fig. \ref{fig:4}(a)-\ref{fig:4}(d) to calculate the right-hand-side of eqn (\ref{eq:meanvel0}) at reduced sampling frequencies $f_0/n < 1500$~Hz, the relationship between $\rho (x|\mathrm{TP})$ and $v_{\mathrm{TP}}(x)$ with $\langle t_{\mathrm{TP}}\rangle$ as a prefactor is apparently satisfied. This can be checked in Figs. \ref{fig:4}(e)-\ref{fig:4}(f) for all the studied particles by comparing the small filled symbols representing the results of the calculation of the right-hand side of eqn (\ref{eq:meanvel0}) with the respective empty symbols depicting the left-hand side of eqn (\ref{eq:meanvel0}), where both quantities were determined at frequencies decreased by $n = 2,5,10,100$. However, we point out that this seeming agreement between both quantities does not guarantee that they portray the actual physical behavior of $\rho (x|\mathrm{TP})$, which must exhibit a maximum around the barrier top where the mean transition path velocity is low, and must significantly diminish close to transition region boundaries, where the particle enters and exits at high speed on average. This behavior can only be retrieved  from the experimental trajectories when recording the particle dynamics at sufficiently high frequency, which ensures that segments that recross
the boundaries of the transition region and that could inexorably be misidentified as transition paths, have little effect on the analysis of their statistical properties. Only under such conditions, eqn (\ref{eq:meanvel0}) provides a physically valid equivalence between $v_{\mathrm{TP}}(x)$ and $\rho (x|\mathrm{TP})$ that quantifies the spatial features of the transition path dynamics of the bead over the barrier.

\begin{figure*}[htb]
 \centering
\includegraphics[width=1.\columnwidth]{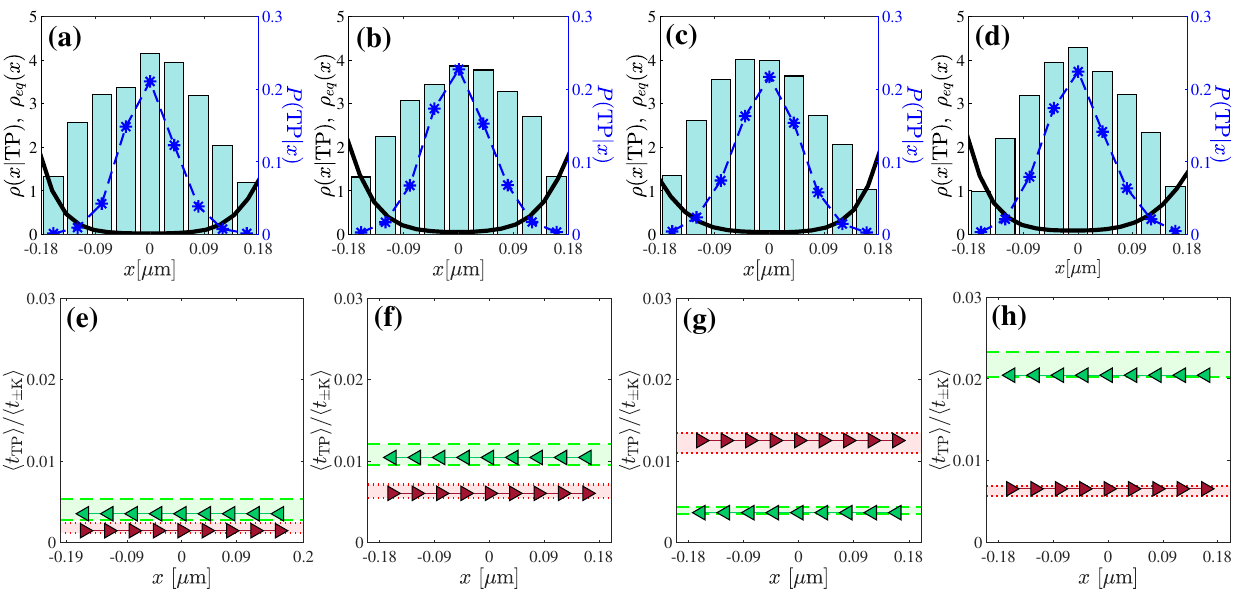}
\caption{(a)-(d) Probability density function of the bead position along transition paths from $-x_0$ to $+x_0$, $\rho (x|\mathrm{TP})$ (vertical bars), equilibrium Boltzmann distribution of the particle position inside the transition region, $\rho_{eq} (x)$ (black solid lines), and probability that at position $x$ in the transition region the particle is on a transition path from $-x_0$ to $+x_0$, $P(\mathrm{TP}|x)$ ($\ast$), all determined from the experimental trajectories of the four analyzed particles.  The dashed lines are just guides to the eye. (e)-(h) Ratio on the right-hand side of eqn (\ref{eq:ratiotime}) that is calculated using the experimental profiles of $P(\mathrm{TP}|x)$, $\rho_{eq} (x)$, and $\rho (x|\mathrm{TP})$, as well as the occupancy probabilities in each well, $P_{\pm}$, for the four analyzed particles undergoing thermally activated transitions from W$_-$ to W$_+$ ($\triangleright$) and from W$_+$ to W$_-$ ($\triangleleft$). The shaded areas enclosed by the dotted and dashed lines represent the experimental values with their respective error bars of the ratio of time-scales on the left-hand side of eqn (\ref{eq:ratiotime}) for transitions paths of the bead taking place from $-x_0$ to $+x_0$ and from $+x_0$ to $-x_0$, respectively.}
 \label{fig:5}
\end{figure*}

Finally, in Fig. \ref{fig:5} we examine the validity of the non-trivial relationship provided by eqn (\ref{eq:ratiotime}), which links the relative magnitude of the two characteristic time scales of the barrier crossing process with the ratio of probabilities and probability densities on its right-hand side. Here, we analyze the experimental data at the original sampling frequency $f_0 = 1500$~Hz, at which we have previously checked that no artifacts are induced on the calculation of the various quantities characterizing the transition path dynamics of the bead. Accordingly, in Figs. \ref{fig:5}(a)-\ref{fig:5}(d) we trace the experimental profiles inside the transition region of the probability density function of the particle position along transition paths from $-x_0$ to $+x_0$, $\rho (x|\mathrm{TP})$, the Boltzmann distribution of the particle position, $\rho_{eq} (x)$, and the probability that at position $x$ the particle is on a transition path from $-x_0$ to $+x_0$, $P(\mathrm{TP}|x)$. The latter is simply determined by counting the fraction of points belonging only to transition paths from $-x_0$ to $+x_0$ that lie within the interval $[x-\Delta x/2, x + \Delta/2]$ with respect to the total number of points found in that region along a full trajectory, including transition paths from $+x_0$ to $-x_0$ as well as trajectory segments that enter the transition region but fail to reach the opposite boundary. Interestingly, we find that $\rho (x|\mathrm{TP})$ and $P(\mathrm{TP}|x)$ are both non-monotonic functions of the coordinate $x$ with a maximum at $x=0$, where the peak of the latter is much sharper than that of former, and  significantly differ from the shape of $\rho_{eq} (x)$, which displays instead a minimum at $x=0$. Moreover, we also compute the occupancy probability of the particle in the well W$_-$, $P_{-}$, by counting the fraction of data points along the complete trajectory over 1 hour that fulfill the condition $x < -x_0$. Then, by use the ensuing values of $P_-$ as well as the corresponding experimental profiles of $\rho_{eq} (x)$,  $\rho (x|\mathrm{TP})$ and $P(\mathrm{TP}|x)$, we calculate the right-hand side of eqn (\ref{eq:ratiotime}) at each discrete location inside the transition region for transition paths going from $-x_0$ to $+x_0$. The results are shown in Figs. \ref{fig:5}(e)-\ref{fig:5}(h) as triangles pointing to the right for all the particles studied in the experiments. A similar analysis of the corresponding probabilities and probability densities for transition paths taking place from $+x_0$ to $-x_0$ leads to the results depicted as triangles pointing to the left in Figs. \ref{fig:5}(e)-\ref{fig:5}(h). Notwithstanding the strongly non-monotonic dependence of $\rho_{eq} (x)$,  $\rho (x|\mathrm{TP})$ and $P(\mathrm{TP}|x)$ on $x$ that is demonstrated in Figs. \ref{fig:5}(a)-\ref{fig:5}(d), we find that the right-hand side of eqn (\ref{eq:ratiotime}) becomes constant across the transition region. Moreover, we directly compute the ratio ${\langle t_{\mathrm{TP}}\rangle}/{\langle t_{\pm \mathrm{K}} \rangle}$ by using the measured values of the mean transition path time and the mean escape times from each potential well with their respective error bars, which are drawn as shaded areas in Figs \ref{fig:5}(a)-\ref{fig:5}(d). Remarkably, we find that the experimental values of this ratio of time-scales agree quantitatively with the right-hand side of eqn (\ref{eq:ratiotime}) for transitions paths occurring in both directions and for all analyzed particles, thereby verifying the underlying relationship between the different pieces of information that describe their thermally activated transitions across the bistable energy landscapes.

\section{Conclusions}\label{sect:conc}

In this work, we have experimentally examined the behavior of the average velocity of the transition paths performed by a submicron-sized bead moving a viscous liquid when successfully surmounting an energy barrier between two optical potential wells. We have verified that the measured velocity profiles exhibit quantitative agreement with those predicted by a diffusive model for the motion of a particle over a parabolic barrier provided that the transition paths are sampled with a time resolution at least two orders of magnitude shorter than their mean duration. This guarantees that the false identification of segments recrossing the boundaries of the transition region as parts of the transitions paths, which inexorably occurs in experiments, has a negligible effect on the determination of their statistical properties. Otherwise, artifacts in the behavior of the mean transition path velocity become evident even when the transition paths followed by the particle over the barrier and their corresponding mean duration are seemingly well detected. Therefore, our findings underline the importance of the sampling rate in the determination of the detailed dynamical features of transition paths in thermally activated processes of small-scaled systems. Furthermore, thanks to the experimental spatio-temporal resolution, we have been able to quantitatively verify a theoretical expression for the mean transition path velocity in terms of the probability density of the particle position along transition paths. In turn, this has also allowed us to experimentally validate an underlying relationship between the mean escape times from the potential wells and the mean transition path times that involves equilibrium and transition path probability distributions in a nontrivial manner. 

In recent years there has been a growing interest in probing numerous aspects of Kramers' escape problem in diverse systems ranging from colloidal particles to biomolecules due to the advancements of experimental mesoscopic techniques during the last decades. Therefore, the results presented here provide valuable insights into the spatio-temporal characterization of barrier crossing processes of diffusive systems beyond the mere determination of transition rates, which could find applications in the analysis and interpretation of thermally activated transitions undergone by more complex systems in contact with a heat reservoir. Investigating the properties of transition paths across barriers of systems that are intrinsically in non-equilibrium states, \textit{e.g.} colloidal particles subject to active forces \cite{militaru2021,paneru2023}, could also be of interest from a fundamental viewpoint as well as for practical applications and will be the subject of further research.

\section*{Author Contributions}

B. R. Ferrer: conceptualization, formal analysis, investigation, methodology, software, validation, visualization. J. R. Gomez-Solano: conceptualization, formal analysis, funding acquisition, investigation, methodology, project administration, resources, software, supervision, validation, visualization, writing – original draft.

\section*{Conflicts of interest}
There are no conflicts to declare.

\section*{Acknowledgements}
We acknowledge support from DGAPA-UNAM PAPIIT Grant No. IA104922.

\end{document}